# How industrial clusters influence the growth of the regional GDP: A new spatial-approach

Vahidin Jeleskovic*   and Steffen Löber**

## Abstract

In this paper, we employ spatial econometric methods to analyze panel data from German NUTS 3 regions. Our goal is to gain a deeper understanding of the significance and interdependence of industry clusters in shaping the dynamics of GDP. To achieve a more nuanced spatial differentiation, we introduce indicator matrices for each industry sector which allows for extending the spatial Durbin model to a new version of it. This approach is essential due to both the economic importance of these sectors and the potential issue of omitted variables. Failing to account for industry sectors can lead to omitted variable bias and estimation problems. To assess the effects of the major industry sectors, we incorporate eight distinct branches of industry into our analysis. According to prevailing economic theory, these clusters should have a positive impact on the regions they are associated with. Our findings indeed reveal highly significant impacts, which can be either positive or negative, of specific sectors on local GDP growth. Spatially, we observe that direct and indirect effects can exhibit opposite signs, indicative of heightened competitiveness within and between industry sectors. Therefore, we recommend that industry sectors should be taken into consideration when conducting spatial analysis of GDP. Doing so allows for a more comprehensive understanding of the economic dynamics at play.

* Humboldt Universität zu Berlin, Email: vahidin.jeleskovic@hu-berlin.de
** Independent researcher

# 1. Introduction

The phenomenon of clustering geographic units in some way is much older than the scientific horizon of this topic in which we want to address the question of the meaningfulness of industry clusters for the economic development of regions. Already Marshall (1961) indicated the great importance of clustering in economics. This was discussed in much higher profundity by Porter (1990) who introduced his famous diamond model for these purposes. For starting an analysis about (industry) clusters, one must at first be aware of correct identification of clusters. Thus, great effort has been done to develop methods for the optimal identification of certain clusters. Examples for that are given e.g. by

The attractiveness of our paper lies in the new way in which one can model and estimate the effects of the of the industry clusters on regional output within a standard spatial econometric model. Hence, we put the focus on introducing a new method for the analysis of industry clusters mostly from the econometric point of view.

For these purposes, we us data from the NUTS 3 regions in Germany for a period of 11 years. We begin by introducing important theoretical basis and lay the foundation of our paper by explaining the cluster theory. Moreover, the empirical methods are presented. The first step is, therefore, a short introduction to panel econometrics and the presentation of the two most important models, the fixed-effects model and the random-effects model. This section is followed by an explanation of the spatial models. The combination of panel econometrics with spatial models makes the paper at hand unique and allows more far-reaching interpretations and estimates than the isolated method.

All variables and clusters evaluated are presented and explained. In order to provide a differentiated view of diverse clusters and sectors, we will look at clusters from eight different industries. In this way it will be possible to assess both, the basic impact of clusters in general and in particular per sector. To this end, we complement the consideration of direct and indirect effects, which will allow a more detailed interpretation of the impact of different sectors. The differentiation of the different clusters is also useful and necessary from a methodological point of view. As will be shown, the distribution of the different clusters varies greatly from sector to sector. One can observe an accumulation in West and South Germany, but here the distribution is very heterogeneous. The heterogeneous and specific distribution of the clusters suggests that the sectors have a systematic influence on the region. In order to reduce problems of endogeneity, it therefore makes sense to differentiate according to the different clusters and their specific effects. Finally, the presentation and interpretation of the estimates and a conclusion are given. This is intended to provide an outlook on future research projects and potential. Through these studies, we hope to better understand the effects of clusters and their impact on regional development. Aristotle already noted: "The totality is not, as it were, a mere heap, but the whole is something besides the parts".

Our investigations will show whether this can be transferred to the concept of clusters.

# 2. Method

## 2.1 Cluster

Already in early times, scientists observed an attraction of certain areas for certain industries. Firstly the British economist Alfred Marshall described this settlement of certain industries in certain regions as „localization of industries" in his book „The Principles of Economics" in the 19th century. Based on his observations, he identified specific main reasons for that localization. The material conditions of the area seem to be of great importance.

Essentially, he highlights the existing properties of the climate and the soil. In addition, he points out that the existence of sources of raw materials and easy accessibility to the region are also important aspects. These important regional characteristics are called „chief causes" (Marshall, 1961, p. 243). Today, based on Marshall's remarks, the US economist Michael E. Porter is the most important researcher in the field of cluster theory. He shaped the idea of clusters in his book „The Competitive Advantage of Nations". He identified certain determinants influencing the innovative strength and competitiveness of countries and firms. He combined these determinants in a dynamic system, called „Porters diamond" (Porter, 1990, p. 72).

### 2.1.1 Porters diamond model

This dynamic system provides a framework for the establishment and growth of national companies (Porter, 1990, p. 72). The determinant „Factor Conditions" includes production factors, necessary for competition (Porter, 1990, pp. 73-75). However, he extended the traditional factors labour, land and capital and divided these into „human resources", „physical resources", „capital resources", he added „knowledge resources" and „Infrastructure". The second determinant „Demand Conditions" describes the domestic demand for goods and services.

This factor is very important for growth and innovative capability of the domestic companies. He identified three characteristics, the structure of the domestic competition, size and growth of the domestic demand and as well as mechanisms for transferring the preferences of a nation's own people to foreign markets.. Furthermore, he observed that the quality of the domestic demand is much more important than the quantity (Porter, 1990, p. 86). The third determinant he identified, is made up by the „Related and Supporting Industries". With this determinant he takes into account the existence of internationally competitive supplier industries in the nation (Porter, 1990, p. 71). The determinant includes the structure of the supplier industries, for example the precursors influence the innovations of a company. As important as the suppliers are related industries as they can pool their research activities which might affect the innovative capability and competitiveness (Porter, 1990, p. 105). As the last determinant of the basic diamond model, Porter identified „Firm Strategy, Structure, and Rivalry". That factor includes the conditions how to start a business, how to lead and organize it as well as the structure of the domestic competition. Porter claimed competition as extremely important for innovative activities of a company and its growth. He observed that pressure and friction give international companies impulses for their development and their ability to innovate (Porter, 1990, pp. 107-117). Porters diamond is a dynamic and interdependent system of determinates that influence each other.

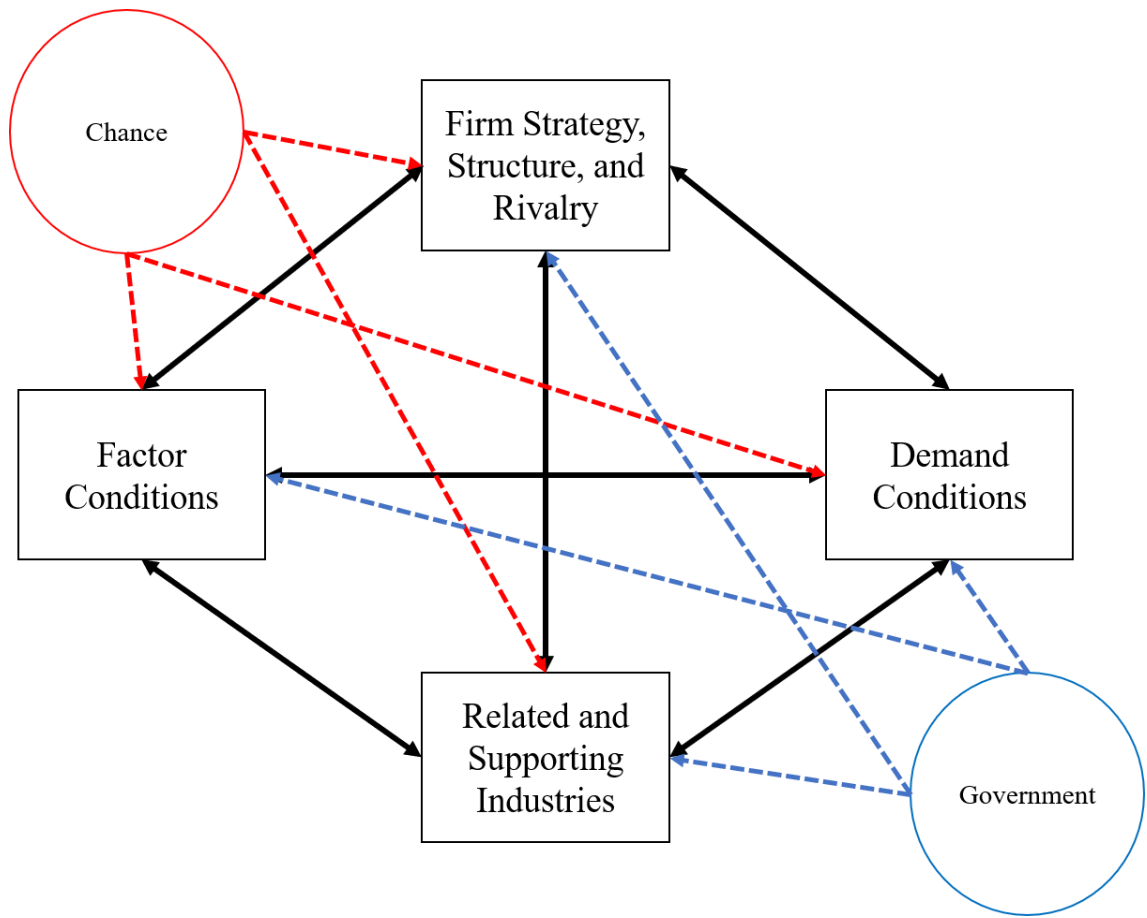

*Figure 1: Diamond model based on Porter (1990, p. 127)*

In addition, the extended model (shown in Figure 1) includes two further factors, which also influence the four basic determinants. On the one hand he added the „Government" and on the other „Chance". Both influence every other factor.

Chance encompasses every aspect that the actors cannot directly influence, for example political developments or emerging basic innovations. The government also influences each of the determinants, for example by economic policy, which naturally affects all factors (Porter, 1990, p. 73).

2.1.2 Concept of cluster

Based on his diamond model, Porter describes clusters as a „geographic concentrations of interconnected companies, specialized suppliers, service providers, firms in related industries, and associated institutions (e.g., universities, standards agencies, trade associations) in a particular field that compete but also cooperate." (2000, p. 15). From that perspective, clusters are accumulations of different interconnected firms and institutions, building a common system. In reference to Porters diamond model, the relations within a cluster are multidimensional and complex. On the one hand, the actors are connected by competition, and on the other hand equally by cooperation. Both are equally important. In that way, they can increase advantages of agglomeration. According to Bröcker & Fritsch (2012, p. 100), a major advantage of agglomeration is that knowledge is exchanged between different actors. Knowledge generated by some actors can be used relatively easily by others and thus, so-called spill-over effects can influence the growth process of regions at a very high level. Already Marshall (1961, p. 245) described the phenomenon and potential of spill-over effects. He maintains that everything begins with an idea, which in turn is taken up and supplemented by others. In addition, other companies are settling in the surrounding area, for example to take over the supply of materials for the new product or its transport. So he pointed out the importance of exchange and these effects on economic development, even if he did not use the term „spill-over" explicitly.
In the empirical part of the paper we will investigate this hypothesis based on empirical analysis. To do so, we will use panel data of the NUTS 3-regions of Germany.

We have included the clusters of eight industries in our model. This will enable us to look at and interpret the effects of different sectors in a differentiated way. The distribution of the clusters within the NUTS 3 is based on a paper of Kosfeld & Titze (2017), in which the clusters of R&D-intensive Industries were identified. The distribution is shown in the following figures sorted by industrial sector:

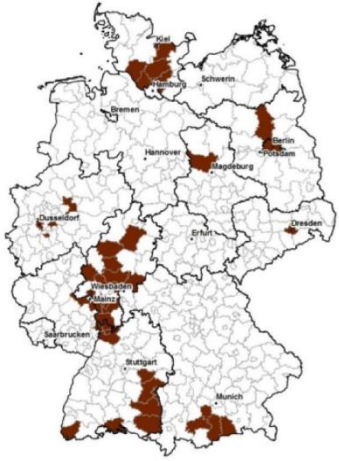 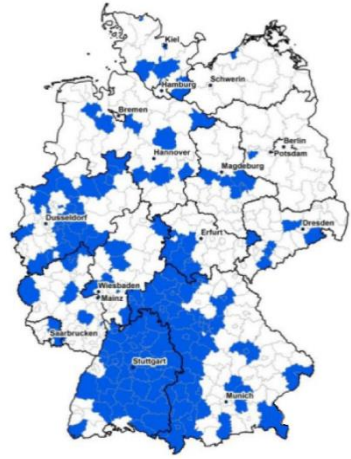

*Figure 2: Chemical clusters*  *Figure 3: Machinery and equipment clusters*

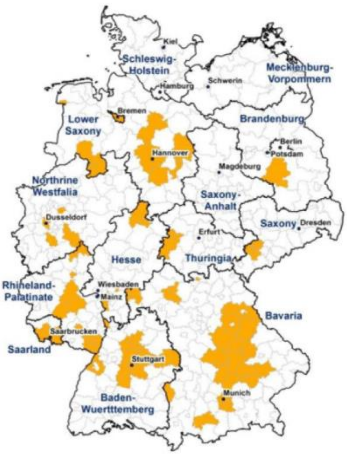 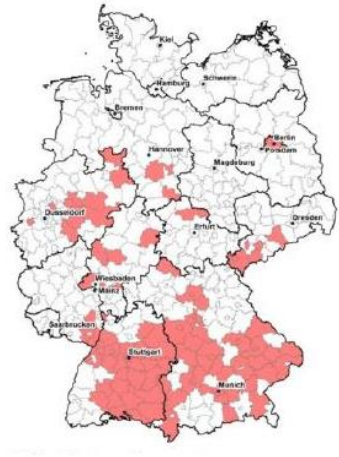

*Figure 4: Automotive clusters*  *Figure 5: Electrical machinery and apparatus clusters*

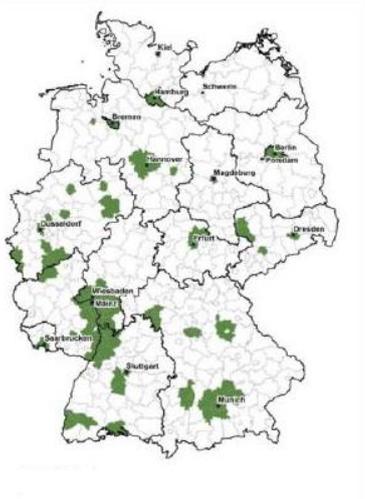 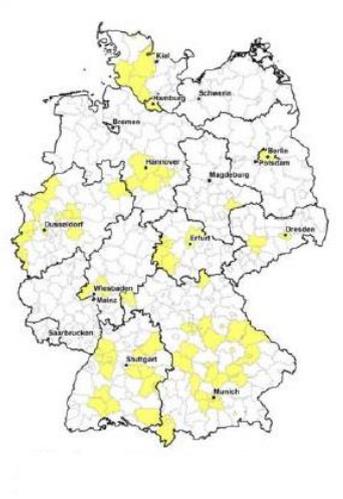

*Figure 6: IT clusters*     *Figure 7: Radio, television, communication equipment and apparatus clusters*

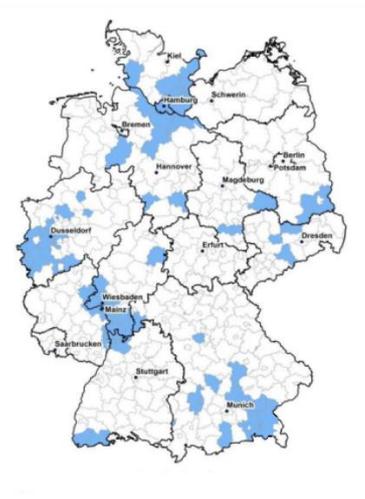 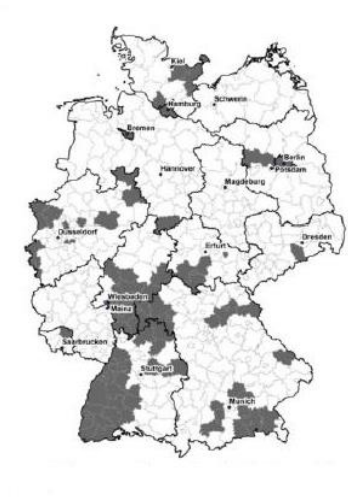

*Figure 8: Pharmaceutical clusters*     *Figure 9: Medical, precision and optical instrument clusters*

As mentioned above, a heterogeneous distribution of the clusters can be identified, which makes a differentiation in the model meaningful. An overview of the exact composition of the clusters and their related industries can be found in the appendix. The different branches of industry within the clusters and their links with each other will still play a role in the interpretation of the empirical results.

## 2.2 Economic theory

Growth concepts, which are often used in the context of clusters, are polarized growth theories. These theories will also form the basis of our economic model and will be briefly explained below. The perspective of these approaches is different from that of neoclassicism or post-Keynesianism. With respect to Werner (2016, p. 16), in the neoclassical approach, production factors and in post-Keynesian theory, demand are the driving force of growth. The classical theories regard imbalances as problematic, whereas polarized growth models deem imbalances as the engine of economic growth. Following Eckey (2008, p. 120), the basis of this approach is the interplay of impulse and resonance. If, for example, production in a region increases due to an innovation, this also affects dependent regions. The higher output increases the demand for intermediate products, which in turn has a positive effect on production. The increased demand means that potential profits attract new players to the market. As a result, new production factors flow into the region, again, increasing productivity due to economies of scale. As productivity increases, so do wages and potential returns on capital. This leads to further inflows of labour and capital. This in turn leads to an increase in productivity and ultimately to a spiral of growth which continues until there is complete spatial concentration of production(Eckey, 2008, p. 121).

The positive effects described are referred to as knock-on effects. Eckey (2008, pp. 120-121) shows that, in contrast, there can also be so-called braking effects. This occurs, for example, when demand for intermediate products cannot be met in one region and resources are poached from other regions. This is known as the suction effect (Eckey, 2008, pp. 120-121). In the following, we will see whether these effects vary from industry sector to industry sector.

## 2.3 Panel models

Based on the panel structure of the data, used in the estimation, special methods for the calculations are necessary. Panel data describe a special type of data with a certain structure. Panel data are characterized by observation of cross-sectional units, for example regions, at different points in time (Arellano, 2003, p. 1). This offers the advantage of a broader database for estimation. Instead of using multiple individuals N that are captured at a time or a person observed at T points, there are NT observations. Due to the wider database, the estimate has a smaller standard error. According to Bauer et al. (2009), the standard error of the estimate is calculated as follows:

$$\sigma_{\hat{\beta}_1} = \sqrt{Var(\hat{\beta}_1)} = \sqrt{\sigma_{\hat{\beta}_1}^2} = \sqrt{\frac{\sigma^2}{\sum_{i=1}^{NT}(x_i-\bar{x})^2}}$$

If NT increases, the denominator also increases, so that the value of the fraction below the square root increases and thus the standard error of the regression $\sigma_{\hat{\beta}_1}$ decreases. Due to the panel structure of the data and the so-called unobserved heterogeneity, the ordinary least squares would lead to inefficient and biased estimates. Therefore, a more specialised method will be used: panel econometrics. Unobserved heterogeneity describes a phenomenon concerning the properties of the observation units. Certain aspects or properties are difficult or impossible to observe and to record, while at the same time they vary across observation units. These aspects are unobserved distinguishing features between the observation units which can, however, influence the dependent variable of the model. A thematic example is infrastructure in a given region at a given time. Infrastructure varies widely from region to region and can be assumed to be constant over time. However, it is difficult to observe and to quantify.

To analyse this type of data anyway, two main models are used, the Fixed Effects Model and the Random Effects Model (Greene, 2012, pp. 385-386).

2.3.1   Fixed effects model

The fixed effects model describes the first advanced model for panel data analysis. In reference to Dreger et al. (2014, p. 261) the following equation is used to show the specific aspects:

$$y_{it} = \alpha_i + \beta' x_{it} + \varepsilon_{it}$$

An important difference compared to the basic regression model is the vector $\alpha_i$. That vector includes the discussed unobserved heterogeneity of a specific observation unit $i$ in the regression model. The assumptions about $\alpha_i$ in turn form the most important difference to the other panel model, the random effects model.

The fixed effects model allows a correlation between the unobserved heterogeneity $\alpha_i$ and one or more of the independent variables $x_{it}$. The assumption about the correlation of the fixed effects model allows:

$$\text{Cov}(\alpha_i, x_{it}) \neq 0$$

To manage that, a so-called within-transformation will be applied. Due to this, all time invariant variables will be eliminated from the model. That method delivers a so-called within estimator, because this estimator only uses the variance within the individuals and not between them. After application of the within-transformation, the model enables unbiased and consistent estimation of $\hat{\beta}$.

### 2.3.2 Random effects model

In contrast to the fixed-effects model, the random effects model does not allow a correlation between the unobserved heterogeneity $\alpha_i$ and one or more of the independent variables $x_{it}$. The following equation in reference to Dreger et al. (2014, pp. 267-272) illustrates the random effects model:

$$y_{it} = x'_{it}\beta + u_{it}$$
$$u_{it} = \alpha_i + \varepsilon_{it}$$

$$\text{with } \text{Cov}(\alpha_i, x_{it}) = 0$$

The structure is very similar to the preceding model but the composition of the error term is different. The error term $u_{it}$ is composed of unobserved heterogeneity $\alpha_i$ and the idiosyncratic error $\varepsilon_{it}$. Furthermore, the following assumption applies:

$$\text{Cov}(\alpha_i, \varepsilon_{it}) = 0$$

Both, the random effects- as well as the fixed effects model are widely common in literature. The difference between the two models can be seen in the quality of the estimators, such as Greene (2012) executes. The random effects model gives unbiased and consistent estimates in the case of no correlation between $\alpha_i$ and one or more exogenous variables. Whereas the fixed effects always deliver consistent estimates, in case of no correlation as well as in case of correlation. Because of elimination of the individual effects before the estimation, the fixed effects estimation is always consistent.

The well-known Hausman-test (Hausman, 1975) is performed to test FE against RE, where $H_0: h = 0$ is tested with:

$$h = d'[var(d)]^{-1}d, \quad d = \hat{\beta}_{FE} - \hat{\beta}_{RE} \quad and$$
$$var(d) = \hat{\sigma}_{RE}^2(X^{\bullet\prime}X^{\bullet})^{-1} - \hat{\sigma}_{FE}^2(X^{*\prime}X^{*})^{-1}$$

$h$ is chi-squared distributed with $K$ degrees of freedom equal the number of explanatory variables excluding the intercept. If one can reject $H_0$, FE may be preferred.

## 2.4   Spatial models

Besides the panel models, spatial models are important method applied. These models consider a phenomenon called spatial dependence or spatial autocorrelation. This is shown by the fact that observed values of one region depend on the values of other regions (LeSage & Pace, 2009).

Due to the spatial dependence between the regions, the ordinary least square leads to inefficient and biased estimates. Therefore, spatial effects must be included in the regression. There are certain methods that consider the effect of spatial autocorrelation. The common factor of all these models is to include a spatial weight matrix $W$ in the estimation. For example, the matrix can be displayed binary and based on neighbourly proximity. In reference to Schulze (1993/94, pp. 60-61) the following presentation could be chosen:

$$W_{ij} = \begin{cases} 1, & if\ i\ is\ a\ neighbour\ of\ j \\ 0, & other \end{cases}$$

Another way to construct the spatial weight matrix is to design it based on the distance between the regions. This method is used in this paper. The factor via which distance is included in the regression is:

$$W_{ij} = \frac{1}{d_{ij}^{\varrho}}$$

where $d_{ij}$ represents the distance from region i to region j and for $i = j$: $W_{jj} = W_{ii} = 0$. In most practical applications, it is set $\varrho = 1$ as well as in our analysis (Duncan, et al., 2017; Earnest, et al., 2007). Because of this, $W$ is a symmetric 402 x 402 matrix with 0 on the diagonal. The distance is placed in the denominator, so $W$ decreases if the distance increases. Thus, the increase in distance in the econometric model can be considered as a decreasing weight. In order to cope with this consideration of spatial dependence, some models exist. According to LeSage & Pace (2009, pp. 25-33), the most common models and their data generation processes are presented below, starting with the spatial lag model (SAR), which includes a spatial lag of the dependent variable:

$$y = \alpha \iota_n + \rho W y + \beta' X + \varepsilon$$
$$y = (I_n + \rho W)^{-1}(\alpha \iota_n + \beta' X) + (I_n + \rho W)^{-1} \varepsilon$$
$$\varepsilon \sim N(0, \sigma^2 I_n)$$

The spatial error model (SEM), including a spatial lag in the error term:

$$y = \alpha \iota_n + \beta' X + u$$
$$u = \rho W u + \varepsilon$$
$$\varepsilon \sim N(0, \sigma^2 I_n)$$

The spatial autoregressive combined model (SAC), that includes a spatial lag of the dependent variable as well as in the error term:

$$y = \alpha \iota_n + \rho W_1 y + \beta' X + u$$
$$u = \theta W_2 u + \varepsilon$$

$$y = (I_n + \rho W_1)^{-1}(\alpha \iota_n + \beta' X) + (I_n + \rho W_1)^{-1}(I_n + \theta W_2)^{-1} \varepsilon$$
$$\varepsilon \sim N(0, \sigma^2 I_n)$$

But to analyse the data, we will use a more suitable model, the so-called Spatial Durbin model (SDM) with an extansion to the SDM-C which is described in more detail in the next section.

### 2.4.1 Spatial Durbin model

The Spatial Durbin model is highly recommended in the literature, because it delivers unbiased estimates. LeSage & Pace (2009, p. 158) already showed, that the SDM is the only model, which produces unbiased estimates for all data generating processes discussed in the preceding section. The SDM can also manage the case of omitted variables, which is often the case when dealing with spatial data. The model also produces correct standard errors or t-values of the estimated coefficients. Even if the actual data generating process shows the characteristics of a spatial error model. This is because the spatial error model is a special case of SDM. In addition, the model does not contain any limitations regarding the dimension of spatial spill over effects (Elhorst, 2010).The spatial dependency is considered both in the dependent variable and in the independent variables. The following equation presents this model in reference to LeSage & Pace (2009, p. 32):

$$y = \alpha \iota_n + \rho W y + \beta' X + W X \gamma + \varepsilon$$
$$y = (I_n + \rho W)^{-1}(\alpha \iota_n + \beta' X + W X \gamma + \varepsilon)$$
$$\varepsilon \sim N(0, \sigma^2 I_n)$$

The factor $\alpha \iota_n$ is a term consisting of the constant $\alpha$ and the nx1 - vector $\iota_n$ consisting of ones. The error term is assumed to be normally distributed with an expected value of zero and homoscedastic. In the model, $\rho$ represents the strength of the spatially lagged dependent variable y. The variable $\gamma$ represents the strength of the spatially lagged explanatory variables X. The transposed vector $\beta'$ quantifies the impact of the explanatory variables to the dependent variable.

For the estimation, methods of panel econometrics are combined with spatial methods.

The biggest advantage of spatial panels is the consideration of spatial and time specific effects. These are effects that vary from region to region, are time-invariant and difficult to observe or measure. If these effects are not considered in the estimates, this can lead to distorted estimates (Elhorst, 2017).

In our analysis, we assume that the primary and most substantial spatial effects in the economy are attributed to industry clusters. Consequently, any spill-over effects outside these industry clusters are considered as random noise. To incorporate this assumption, we use a predetermined identification matrix $C_v$ which determines whether a spatial unit $i$ belongs to the cluster $v$.

$$Clustervariable_{ic} := c_v = \begin{cases} 1, & \text{if their is cluster } v \text{ in that region } i \\ 0, & \text{if their is no cluster } v \text{ in that region } i \end{cases}$$

For regions outside industry clusters, we assume that spill-over effects have an expected value of zero and a constant variance, as per our assumption. With these considerations, we present the Spatial Durbin Model with modeling of clusters (SDM-C) as follows:

$$y = \alpha \iota_n + \rho W y + \beta' X + (W \odot C) X \gamma + \xi$$

where $\odot$ is the Hadamard product. It's important to note that the modeling of cluster effects is incorporated within the exogenous variables, and this modification does not alter the fundamental properties of the SDM.

## 2.5 Direct, Indirect and Total Effects

Using the SDM offers further advantages. LeSage & Pace (LeSage & Pace, 2009) show that the interpretation possibilities of models with spatially delayed dependent or independent variables are more complicated but also more rewarding. This model makes it possible to interpret so-called direct and indirect effects. Direct effects include changes in the dependent variable of region $i$ when an independent variable of the same region $i$ is changed. Indirect effects, on the other hand, include the effect of a change in the independent variable of region $i$ on the dependent variable of region $j$. The following applies: $i \neq j$.

In the following, based on LeSage & Pace (2009) and Pijnenburg & Kholodilin (2012), we see how these summary measures are calculated.[3] We use the so-called spatial multiplier for the reformulation of the equation:

$$(I_n + \rho W)^{-1} = I_n + \rho W + \rho^2 W^2 + \rho^3 W^3 + \cdots$$

$$(I_n + \rho W) y = \alpha \iota_n + \beta X + W X \gamma + \varepsilon$$
$$y = (I_n + \rho W)^{-1} (\alpha \iota_n + \beta X + W X \gamma + \varepsilon)$$

Which we can transform into:

$$(I_n + \rho W)^{-1} (\beta' X + W X \gamma) = \sum_{r=1}^{k} (I_n + \rho W)^{-1} (I_n \beta_r + W \gamma_r) x_r$$
$$= \sum_{r=1}^{k} S_r(W) x_r$$

---

[3] For other details, see Elhorst (2010)

In this formulation the index *r* represents the exogenous variable.

Looking at the data generating process (e.g. in Pijnenburg & Kholodilin, 2012, p. 9), the following result appears after the derivation from $y_i$ to $x_{jr}$:

$$\frac{\partial y_i}{\partial x_{jr}} = S_r(W)_{ij}$$

The derivation does not correspond to the coefficient $\beta_r$. However, it can be shown that a change in the exogenous variables *r* of region *j* can have an influence on the dependent variable of region *i*.

If $y_i$ is now derived after $x_{ir}$, the following picture appears:

$$\frac{\partial y_i}{\partial x_{ir}} = S_r(W)_{ii}$$

Here again, the derivation does not correspond to the coefficient $\beta_r$.

If one derives according to *i*th region, one obtains a term $S_r(W)_{ii}$. This term describes the influence on $y_i$ when there is a change in $x_{ir}$. This results in so-called *feedback loops*. The region *i* influences the region *j* and vice versa. Also included are effects that affect several regions. The region *i* influences the region *j* which in turn influences the region *k*. Starting from region *k* the effect goes back to region *I* (LeSage & Pace, 2009).

In order to derive the direct and indirect effects, the matrix $S_r(W)$ serves as a basis for calculation. Following Pijnenburg & Kholodilin (2012, pp. 9-10) the average direct effect results from the calculation of the trace of the matrix:

$$\bar{M}(r)_{direct} = \frac{1}{n} tr(S_r(W))$$

To calculate the average total effect, the average of all derivatives from $y_i$ to $x_{jr}$ is established for all regions.

$$\bar{M}(r)_{total} = \frac{1}{n} \iota'_n (S_r(W)) \iota_n$$

By forming the difference of $\bar{M}(r)_{direct}$ and $\bar{M}(r)_{total}$ the average indirect effect can now be derived:

$$\bar{M}(r)_{indirekt} = \bar{M}(r)_{total} - \bar{M}(r)_{direct}$$

Understanding these summary measures will allow the results of the estimates to be examined in greater depth in the following sections.

## 2.6 Economic model

A set of variables was used to carry out the estimation. We have used these factors in the form of the following production function:

$$Y = F(I, P, S, L)$$

The dependent variable $Y$ is the $GDP_{it}$ (gross domestic product) for every region $i$ at time $t$ of the German NUTS 3 regions[4]. The data set includes observations for 402 German cities and counties. For every region, 11 observations were used and values for years 2002 - 2012. The $GDP$ is used to measure the economic output of a certain region.

As independent variables, four values were included. Firstly, $Income_{it}$ ($I$) was implied in the equation, which represents the disposable income per inhabitant in a region $i$ at time $t$. The second independent variable contains the $Patents$ ($P$) or more precise the patent applications to the EPO ($Patents_{it}$) in a region $i$ at time $t$. The variable $Size_{it}$ ($S$) include the average company size in a certain region $i$ at time $t$. The company size was calculated in the first step as the average count of employees per company. As last explanatory variable we included the economic factor $Labour_{it}$ ($L$) in the estimation. That variable includes the working population in thousands in a region $i$ at time $t$.

We were particularly interested in measuring the impact of clusters on regional $GDP$. Based on the shown differentiation according to industries, we include a matrix for each individual cluster in the regression. These matrices were constructed in advance on the basis of the spatial distributions and are designed as dummy variables with a binary structure. In reference to Bauer et al. (2009, pp. 82-83) that type of variable is characterized by the following properties:

---

[4] NUTS mean the abbriviation for french „Nomenclature des unités territoriales statistiques"

$$Clustervariable_{ic} := d_i = \begin{cases} 1, & \text{if their is cluster c in that region i} \\ 0, & \text{if their is no cluster c in that region i} \end{cases}$$

In contrast to the other variables, we suggest the time invariance of the clusters for the observed period. The cluster varies between the observation units but not over time. This assumption is based on the fact that clusters are not a short-term phenomenon in a region but need both time for growth and time to disappear. Therefore, we regard them as time-invariant. To consider the effects of the financial crisis, we included additional a *Dummy* variable for the year 2009.

As already mentioned, the SDM is used for the estimation. The following equation will therefore be the basis for this paper:

$$\begin{aligned} \ln GDP_{it} = {}& \alpha_i + \rho W * \ln GDP_{it} \\ & + \beta_1 * \ln Income_{it} + \beta_2 * \ln Labour_{it} + \beta_3 * \ln Patents_{it} + \beta_4 * \ln Size_{it} \\ & + \beta_5 * Dummy \\ & + \gamma_{1c} * W \odot Clustervariable_{ic} * \ln Income_{it} \\ & + \gamma_{2c} * W \odot Clustervariable_{ic} * \ln Labour_{it} \\ & + \gamma_{3c} * W \odot Clustervariable_{ic} * \ln Patents_{it} \\ & + \gamma_{4c} * W \odot Clustervariable_{ic} * \ln Size_{it} \\ & + \xi_{it}, \end{aligned}$$

where $\odot$ Hadamard-product is. The time constant vector $\alpha_i$ contains the unobserved heterogeneity of region *i*. $W * \ln GDP_{it}$ is the spatial lag of the regional *GDP*, *WX* includes the spatial lag of the explanatory variables *X={ln Income, ln Labour, ln Patents, ln Size}*. Finally, $\xi_{it}$ is considered as error term in the regression which captures all unobservable factors influencing $GDP_{it}$. Furthermore, these are logarithmic values, which allows a more intuitive interpretation of the results as elasticities.

We point our here that we used the Hausman test as well for this model specification. The result of the Hausman test comes out by 24 degrees of freedom with the test statistic to be 487.41. This implies rejection of random effects model in favour of fixed effects model at 1% significance level. As Baltagi (2005) and Elhorst (2003) show, the Hausman-test can also be used for spatial panel models to test FE against RE.

## 3. Empirical results

In Table 1, only significant coefficients are presented, with t-statistics reported in parentheses throughout the rest of the paper.

| | | Income | Labour | Patents | Size | Dummy | rho ($\rho$) |
|---|---|---|---|---|---|---|---|
| $R^2$ | 0.981 | | | | | | |
| Adj. $R^2$ | 0.9797 | | | | | | |
| | | 0.58 (37.34097) | 1.02 (241.7505) | 0.02 (9.383073) | 0.23 (28.70612) | -0.03 (-4.63188) | 0.003 (30.60779) |
| Cluster | Automotive | -0.53 (-6.37913) | -0.08 (-5.367449) | 0.08 (9.383073) | -0.13 (28.70612) | | |
| | Machinery and equipment | 0.27 (3.53008) | 0.08 (5.17147) | -0.06 (-4.87353) | 0.13 (3.26441) | | |
| | Electrical machinery and apparatus | 0.25 (3.20813) | - | -0.02 (-1.66706) | - | | |
| | Chemical | - | -0.11 (-2.97355) | 0.05 (2.71709) | - | | |
| | Medical, precision and optical instrument | - | - | - | 0.21 (4.03829) | | |
| | Pharmaceutical | - | 0.05 (3.00379) | - | - | | |
| | IT | 0.39 (3.0653) | 0.06 (2.08996) | -0.05 ( -2.08223) | | | |
| | Radio, television, communication equipment and apparatus | - | - | - | -0.32 (-3.72645) | | |

*Table 1: Estimation output*

The value of $R^2$ signifies, that the model can explain 98.1% of the variance within the data (or *Adj. $R^2$* still 97.97%). The *rho* as the indicator for the strength of the spatial lag dependence of the GDP is very small but highly significant. If the *GDP* in a dependent region increases by 10 percent, the *GDP* in the own region increases by 0.03 percent. The value indicates, the *GDP* in a certain region has a very small positive impact to the *GDP* of a dependent region. These effects would be neglected, if a model without spatial effects had been used.

As explained, the Dummy variable enables the model to consider the effects of the financial crisis. As result, a highly significant negative impact on the *GDP* could be detected. The explanatory variables without the implied spatial lag are all positive and highly significant.

The *Income,* firstly estimated without spatial effects, has a highly significant positive impact on the local *GDP*. If *Income* in one's own region increases by 10 percent, the *GDP* in one's own region increases by 5,8 percent. For *Patents,* we detected a very small impact on the regional *GDP* whereas the coefficient of *Labour* demonstrates a relatively strong impact. If the number of *Patents* in the own region increases by 10 percent, the *GDP* in the own region increases by 0.2 percent. However, if *Labour* increases by 10 percent in its own region, the *GDP* increases by 10.2 percent in its own region. The production factor *Labour* can therefore be assumed to have a strong influence on regional growth. The last exogenous variable is *Size*. If this factor increases by 10 percent in one's own region, *GDP* in one's own region increases by 2.3 percent. Especially for clusters of the automotive sector, strong negative spatial effects could be discovered. If the I*ncome* increases by 10 percent in the spatially dependent regions with an existing Automotive cluster, the *GDP* decreases by 5.3 percent in the own region. The same applies, to a lesser extent, to *Labour* and *Size*. Only in existing Automotive clusters, *patents* have a weakly positive effect on GDP. With reference to the polarized growth theory presented, the coefficients show a suction effect starting from the automotive clusters on spatially dependent regions. Especially related to the factors *Income*, *Labour* and *Size*. The assumption suggests that an automotive cluster in a dependent region can lead to an outflow of human capital from the corresponding regions. As the FAZ reports,[5] automotive companies was again seen the most popular employers for graduates in 2019. This can lead to many workers leaving their home regions to work in production facilities of the automotive industry. An increase in *Income*, *Labour* and *Size* induces the influx of labour into the cluster. The shortage of these workers in their home regions may lead to a decline in their economic performance. Looking at Machinery and equipment clusters, the following picture emerges: If the *Income* increases by 10 percent in the dependent regions with Machinery and equipment clusters, *GDP* increase by 2.7 percent in the own region. The same positive effect is true for *Labour* and *Size*.

---

[5] https://www.faz.net/aktuell/karriere-hochschule/buero-co/beliebte-branchen-junge-akademiker-reizt-weiterhin-die-autoindustrie-16421282.html

*Patents*, on the other hand, have a weakly negative influence. In contrast to the clusters of the automotive sector, machinery- and equipment clusters seem to have a knock-on effect on spatially dependent regions and thus increase their economic performance. One reason could lie in the products themselves. Especially in comparison to the automotive industry.

Actors in the Machinery and equipment clusters are active, among other things, in the manufacture of machines. These are used later to increase production output. The following example can serve as an illustration: In one cluster, a machine is constructed for the production of a certain product. The customer who buys this machine uses it to increase his own production. The machine from the cluster, and thus the cluster itself, has a positive influence on the performance of the customer and his region.

In the dependent regions with Electrical machinery- and apparatus clusters, the increase in *Income* by 10 percent leads to an increase in *GDP* in the own region by 2.5 percent. For *Patents*, the increase of 10 percent leads to a very small reduction of *GDP* in the own region by 0.2 percent. This very small effect can be observed for all clusters investigated. The number of patents seems to have only a very small influence on the economic performance of a region. We see this both in the coefficient without spatial effects and in the coefficients with spatial effects. The electrical machinery and apparatus clusters also appear to have a positive impact on the economic performance of dependent regions via knock-on effects. The effects found show very similar characteristics as the coefficients of the Machinery and equipment clusters. If the composition of the electrical machinery and apparatus clusters and the overlaps with the Machinery and equipment clusters are taken into account, it can be concluded that both branches of industry have similar effects on regional development.

In the dependent regions with Chemical clusters, the increase of *Labour* by 10 percent leads to a small reduction of *GDP* in the own region by 1.1 percent. These clusters show a negative effect on dependent regions. For Medical-, precision- and optical instrument clusters, only the estimates for *Size* were significant: The results show that in the dependent regions with medical clusters the increase of *Size* by 10 percent leads to an increase of *GDP* in the own region of 2.1 percent. For Medical-, precision- and optical instrument clusters, positive impulse effects can thus be proven. In the dependent regions with Pharmaceutical clusters the increase of *Labour* by 10 percent leads to an average increase of *GDP* in the own region by 0.5 percent.

IT clusters show positive effects for *Labour* and *Income*. In the dependent regions with IT clusters the increase of *Income* by 10 percent leads to an increase of *GDP* in the own region by 3.9 percent. If Labour increases by 10 percent, GDP rises by 0.6 percent. So, IT clusters show a positive effect on spatially dependent regions. The interpretation can be analogous to

Machinery and equipment clusters. With its products and services an IT cluster supports the productivity and economic performance of its customers and thus has a positive effect on the development of the regions. For radio-, television-, communication equipment- and apparatus clusters only significant results for the factor *Size* could be observed: In the dependent regions with a Radio, television, communication equipment and apparatus cluster, the increase of *Size* by 10 percent leads to a reduction of *GDP* in the own region by 3.2 percent. This supports the theory that the produced goods of a cluster affect the dependent regions. For the summary measures, we get the following results:

|  |  |  | direct | indirect | Total |
|---|---|---|---|---|---|
| **Cluster** | **Automotive** | Income | 0.8065 (22.5453) | -1.0331 (-6.3113) | -0.2267 (-1.7106) |
|  |  | Labour | 1.0524 (147.1518) | -0.1391 (-4.4240) | 0.9132 (34.9524) |
|  |  | Patents | -0.0106 (-1.8052) | 0.1508 (5.6285) | 0.1402 (6.5005) |
|  |  | Size | 0.2864 (15.9374) | -0.2580 (-3.2800) | 0.0284 (0.4502) |
|  | **Machinery and equipment** | Income | -0.6426 (-6.1808) | 0.5259 (3.4328) | -0.1167 (-1.1150) |
|  |  | Labour | -0.1163 (-5.1531) | 0.1559 (5.0187) | 0.0395 (4.6113) |
|  |  | Patents | 0.1001 (5.8414) | -0.1123 (-4.9322) | -0.0122 (-0.8817) |
|  |  | Size | -0.1889 (-3.3439) | 0.2590 (3.3327) | 0.0701 (3.2860) |
|  | **Electrical machinery and apparatus** | Income | 0.1630 (1.6837) | 0.5103 (3.3405) | 0.6733 (6.0151) |
|  |  | Patents | 0.0872 (5.2128) | -0.0342 (-1.6387) | 0.0530 (2.3783) |
|  | **Chemical** | Labour | -0.0098 (-0.5158) | -0.2233 (-3.0108) | -0.2331 (-3.8693) |
|  |  | Patents | 0.1101 (2.7263) | 0.1050 (2.9079) | 0.2151 (4.3625) |
|  | **Medical, precision and optical instrument** | Size | 0.1614 (2.0043) | 0.4356 (4.1762) | 0.5970 (5.2664) |

| | Pharmaceutical | Labour | -0.0436<br>(-3.3392) | 0.1179<br>(2.8853) | 0.0743<br>(2.1324) |
| | | Income | -0.2788<br>(-4.2852) | 0.7763<br>(3.1382) | 0.4974<br>(2.5126) |
| | IT | Labour | 0.0273<br>(1.4939) | 0.1127<br>(2.1730) | 0.1400<br>(2.8321) |
| | | Patents | 0.2357<br>(4.4556) | -0.0858<br>(-2.0225) | 0.1499<br>(2.3268) |
| | Radio, television, communication equipment and apparatus | Size | 0.1941<br>(4.4076) | -0.6254<br>(-3.6035) | -0.4313<br>(-3.1590) |

*Table 2: Summary measures*

For most part, the results support the estimates obtained, but others allow for a much more differentiated interpretation. The effects of *Income* vary greatly depending on the cluster. While we were able to demonstrate positive direct effects in the Automotive and Electrical machinery and apparatus cluster, the factor shows negative significant effects in the Machinery and equipment and IT sectors. So, we see that the change in *Income* in region *i* can have different effects on region *i* itself, depending on the cluster. Especially in the Automotive sector, *Income* has positive direct effects for the regions. The rising *Income* thus creates incentives for workers to migrate to the regions of the cluster, which leads to an increase in the economic output of the regions. For the indirect effects we see very strong negative effects especially for clusters in the Automotive sector. Which means that an increase in *Income* in a connected region leads to a decrease in the economic performance of its own region *i*. This can be attributed to the *Income* incentives for the labour force. An increase in *Income* in a connected region attracts labour from other regions. This in turn leads to a decline in the economic performance of the "bleeding" regions. A further reason can be found in the dominance and attractiveness of the Automotive sector for skilled workers. That leads to a strong migration of human capital into the automotive clusters and to a brain drain from other regions. We can therefore actually observe a braking or suction effect here. For IT, Machinery and equipment and Electrical machinery and apparatus clusters, on the other hand, we see positive effects. Rising *Incomes* in the associated regions thus increase the economic performance of their own region *i*. Which suggests that rising incomes in parts of the cluster are also beneficial for the other cluster regions and the actors thus participate in the economic success of the cluster. When we look at the *Labour* factor, we see similar trends as for *Income*. We find strong positive direct effects in the automotive sector in particular. In contrast, however, a negative effect of the variable in regions with Machinery and equipment clusters. Negative indirect effects were observed in the Automotive and Chemical industries, while they were positive in the Machinery and equipment and IT sectors. The values found for *Labour* are to be interpreted analogously to those of the factor *Income*. For the *Patents* factor, we were able to demonstrate positive direct effects for some sectors, but very small negative effects in the regions with Automotive clusters. An increase in *Patents* can be interpreted as an increase in economic efficiency, which has a direct impact on the output of a region. For the indirect effects the picture turns around, while these are positive for the Automotive cluster, they are negative for clusters of Machinery and equipment, Electrical machinery and apparatus and IT. The positive indirect effects can be interpreted as knowledge-spillovers. Looking at the Automotive and Machinery and equipment sectors, an increase in *Patents* in related regions leads to an increase in economic performance. This speaks for close cooperation within the clusters as well as for participation in *Patents*, for example in the form of joint research or innovation cooperation. It is particularly noticeable that the positive effects can be found in the research-intensive sectors of the Automotive and Chemical industries. The negative indirect effects of the *Patents* factor could have to do with knowledge outflows from related regions, i.e. brain drain. For the factor *Size* we could prove positive direct effects for regions with Automotive, Medical, precision and optical instrument and Radio, television, communication equipment and apparatus cluster. However, we found negative direct effects for the Machinery and equipment sector. For the indirect effects, however, we were able to demonstrate positive effects for the Machinery and equipment and Medical, precision and optical instrument sectors. The Automotive and Radio, television, communication equipment and apparatus sector, on the other hand, showed negative indirect effects. The effects of the *Size* factor show the same trends for the clusters as we were able to show for *Income* and *Labour*. For the Automotive sector, large company sizes in the regions show positive effects on the corresponding region. At the same time, the strong negative indirect effects show that large company sizes in connected regions have a negative impact on the economic performance of their own region *i*. This can also be explained by the outflow of important resources to the regions with the production sites. At the same time, we see negative effects of company size in Machinery and equipment clusters. Here it seems to be advantageous to have small operating sites in the own region *i*. However, the own region seems to be positively influenced by the size of the companies in connected regions. With increasing company size, the output of own region also increases. This argues in favour of cross-regional supply and relationship chains, through which connected regions can benefit from large companies in the cluster. A closer look at the direct effects of the various clusters shows that these sometimes deviate very strongly from the coefficients previously calculated for the four independent variables. Hence, differentiating the clusters seems to be of enormous importance, as extreme differences between the various sectors are evident. Moreover, the coefficients that incorporate the spatial aspects of the clusters may seem somewhat too low, especially compared to the estimates of the indirect effects of the various sectors. It is therefore important to note that distinguishing between the different sectors brings a clear gain in knowledge, as we can see enormous differences between the betas of the independent variables compared to their clustered direct effects. Furthermore, it can be stated that although a consideration of the spatial lag in the independent variables allows an assessment of the influence, this influence is obviously underestimated by the coefficients. However, we also see a problem with the combination of positive and negative effects. Due to their mathematical derivation from the direct and indirect effects, the total effects become very small and insignificant for some clusters. The effects cancel each other out, so that a neutralization effect between direct and indirect effects can arise.

## 4. Conclusion

After estimating the effects, the data reveal that clusters have a modest yet highly significant impact on regional GDP. However, we also observe a degree of complexity when it comes to the clusters of different industries and their isolated effects on regional growth. It becomes apparent that the economic impact doesn't solely hinge on the presence of a cluster but also on the specific industry it represents. This observation gains further support when considering both direct and indirect effects.

The analysis reveals significant disparities between the estimated coefficients and summary measures, underscoring the need to accord a high level of importance to the interpretation of the sector under investigation. Within these clusters, we find notable variations, with some exhibiting significantly positive effects while others demonstrate negative effects. To understand these discrepancies, one must delve into the unique characteristics and intricacies of each specific sector.

For instance, we observe notably strong negative effects in the automotive sector, the chemical sector, and the radio, television, communication equipment, and apparatus clusters. These findings suggest that these sectors possess distinct properties that exert a gravitational pull on spatially dependent regions. A deeper examination of direct and indirect effects provides valuable insights. Both spatial lagged variables and indirect effects showcase pronounced negative impacts on the spatially dependent regions.

Nevertheless, the direct effects reveal significant positive impacts, while these negative effects exemplify the presence of negative spatial autocorrelation, a phenomenon frequently observed in spatial structures. Wicht et al. (2019, p. 12) have also detected such effects in their research, indicating that negative spatial autocorrelation is a characteristic feature of specific areas, particularly in larger cities. In these areas, we often observe high unemployment rates within the city itself, contrasting with relatively low unemployment rates in the surrounding regions.

Another contributing factor could be the highly specialized nature of certain sectors and their resource intensity. Tichy (2001, p. 197) describes a bell-shaped development process, particularly applicable to highly specialized clusters. This process typically involves rapid expansion, followed by (over)saturation, eventual stabilization, and, in some cases, contraction. Sternberg et al. (Sternberg et al., 2004, p. 167) also emphasize the potential drawback of excessive specialization, especially when a cluster is exclusively focused on a single industry.

To gain further insights, additional research is imperative. In this paper, we deliberately refrained from utilizing more complex models and approaches, aiming to highlight the potential for future research in this field. The realm of spatial models remains largely unexplored, and even variations in the weight matrix alone offer significant opportunities. Pijnenburg & Kholodilin (2012, pp. 874-875) underscore the influence and importance of weight matrices, as evidenced by their comparison of four types of geographical distance measures. Their results vary significantly depending on the matrix employed.

In addition to the Spatial Durbin Model we have employed, it's worth considering and comparing other models to identify the most suitable one for our dataset's structure. Furthermore, in the realm of panel econometrics, there are alternative models that merit exploration. While we have utilized the basic fixed effects model, the domain of panel data analysis offers a range of models to consider.

Consequently, we find ourselves at the outset of extensive research endeavors that promise to illuminate the significance of clusters in regional development further. Particularly when dealing with sectors characterized by highly heterogeneous economic effects, a more in-depth exploration is warranted in the future. The sectors themselves represent the linchpin of successful regional development, and both the originating and dependent regions stand to benefit if the findings are taken into account.

Nonetheless, our research underscores that clusters possess the potential to exert a positive influence on regional growth as well as some economic risks, making them valuable contributors to successful economic policies, both in Germany and globally.

# References

Arellano, M., 2003. *Panel Data Econometrics. Advanced Texts in Econometrics.* Oxford: Oxford University Press.

Baltagi, B. H., 2005. *Econometric Analysis of Panel Data.* 3. ed. Chichester: John Wiley & Sons, Ltd.

Bauer, T. K., Fertig, M. & Schmidt, C. M., 2009. *Empirische Wirtschaftsforschung. Eine Einführung.* Berlin Heidelberg: Springer.

Bröcker, J. & Fritsch, M., 2012. *Ökonomische Geographie.* München: Vahlen.

Dreger, C., Kosfeld, R. & Eckey, H.-F., 2014. *Ökonometrie. Grundlagen - Methoden - Beispiele.* 5. ed. Wiesbaden: Springer Gabler.

Duncan, E. W., White, N. M. & Mengersen, K., 2017. Spatial smoothing in Bayesian models: a comparison of weights matrix specifications and their impact on inference. *International Journal of Health Geographics*, 16(47).

Earnest, A. et al., 2007. Evaluating the effect of neighbourhood weight matrices on smoothing properties of Conditional Autoregressive (CAR) models. *International Journal of Health Geographics*, 6(54).

Eckey, H.-F., 2008. *Regionalökonomie.* Wiesbaden: Gabler.

Elhorst, J. P., 2003. Specification and Estimation of Spatial Panel Data Models. *International Regional Science Review*, 26(3), pp. 244 - 268.

Elhorst, J. P., 2010. Applied Spatial Econometrics: Raising the Bar. *Spatial Economic Analysis*, 5(1), pp. 9-28.

Elhorst, J. P., 2017. Spatial Panel Data Analysis. In: S. Shekhar, H. Xiong & X. Zhou, eds. *Encyclopedia of GIS.* 2. ed. Cham, Switzerland: Springer International Publishing, pp. 2050-2058.

Greene, W. H., 2012. *Econometric Analysis.* 7. ed. Harlow: Pearson Education Limited.

Hausman, J. A., 1975. An instrumental variables approach to full information estimators for linear and certain nonlinear econometric models. *Econometrica*, Issue 43, pp. 727 - 738.

Kosfeld, R. & Titze, M., 2017. Benchmark Value-added Chains and Regional Clusters in R&D-intensive Industries. *International Regional Science Review,* 40(5), pp. 530-558.

LeSage, J. & Pace, R. K., 2009. *Introduction to Spatial Econometrics.* Boca Raton: CRC Press.

LeSage, J. P. & Fischer, M. M., 2008. Spatial Growth Regressions: Model Specification, Estimation and Interpretation. *Spatial Economic Analysis*, 3(3).

Marshall, A., 1961. *Principles of Economics. An introductory volume.* 8. ed. London: English Language Book Society and Macmillan & Co. Ltd..

Pijnenburg, K. & Kholodilin, K. A., 2012. Do Regions with Entrepreneurial Neighbours Perform Better? A Spatial Econometric Approach for German Regions. *Regional Studies*, 48(5), pp. 866-882.

Porter, M. E., 1990. *The Competitive Advantage of Nations.* New York: The Free Press.

Porter, M. E., 2000. Location, Competition, and Economic Development. Local Clusters in a Global Economy. *Economic Development Quarterly*, Februar, 14(1), pp. 15-34.

Schulze, P. M., 1993/94. Zur Messung räumlicher Autokorrelation. In: *Jahrbuch der Regionalwissenschaft.* s.l.:s.n., pp. 57-78.

Sternberg, R., Kiese, M. & Schätzl, L., 2004. Clusteransätze in der regionalen Wirtschaftsförderung. Theoretische Überlegungen und empirische Beispiele aus Wolfsburg und Hannover. *Zeitschrift für Wirtschaftsgeographie*, 3-4(48), pp. 164-181.

Tichy, G., 2001. Regionale Kompetenzzyklen - Zur Bedeutung von Produktlebenszyklus- und Clusteransätzen im regionalen Kontext. *Zeitschrift für Wirtschaftsgeographie*, 3-4(45), pp. 181-201.

Werner, A., 2016. *Wachstumsdeterminanten in Deutschland. Quantilsregression und räumlich ökonometrische Analyse regionaler und sektoraler Unterschiede.* Wiesbaden: Springer Gabler.

Wicht, A., Kropp, P. & Schwengler, B., 2019. Are functional regions more homogeneous than administrative regions? A test using hierarchical linear models. *Papers in Regional Science*, 12 Juli, pp. 1-30.

# Appendix

Annex 1

| Cluster templates[a] | Related industries[b] |
|---|---|
| Automotive cluster (34) | 25.2, 28, 31 |
| Chemical cluster (24\24.4) | 17, 19, 20, 21.2, 22.2–22.3, 24.4, 25.1, 25.2, 26.1, 26.2–26.8, 27.4, 27.5, 36 |
| Pharmaceutical cluster (24.4) | 24\24.4 |
| Machinery and equipment cluster (29) | 25.1, 25.2, 26.1, 26.2–26.8. 27.1–27.3, 27.5, 28, 31, 35, 36 |
| IT cluster (30 and 72) | 28, 64, 73 |
| Electrical machinery and apparatus clusters (31) | 28, 29, 33, 34, 35 |
| Radio, television, communication equipment and apparatus clusters (32) | 28 |
| Medical, precision and optical instruments clusters (33) | 25.2, 28, 31 |

[a]Numbers in parentheses represent the sector codes
[b]The description of R&D-intensive industries and their related sectors appears in Annex 2

Source: Kosfeld & Titze, 2017, p. 544.

Annex 2

| Code | Sector |
|---|---|
| 17 | Manufacture of textiles |
| 19 | Manufacture of leather and leather products |
| 20 | Manufacture of wood and wood products |
| 21.1 | Manufacture of pulp, paper and paperboard |
| 21.2 | Manufacture of articles of paper and paperboard |
| 22.2–22.3 | Printing and service activities related to printing; reproduction of recorded media |
| 24\24.4 | Manufacture of chemicals and chemical products |
| 24.4 | Manufacture of pharmaceuticals, medical chemicals and botanical products |
| 25.1 | Manufacture of rubber products |
| 25.2 | Manufacture of plastic products |
| 26.1 | Manufacture of glass and glass products |
| 26\26.1 | Manufacture of other non-metallic mineral products without glass and glass products |
| 27.1 | Manufacture of basic iron and steel and of ferro-alloys |
| 27.3 | Manufacture of tubes; Other first processing of iron and steel |
| 27.4 | Manufacture of basic precious and non-ferrous metals |
| 27.5 | Casting of metals |
| 28 | Manufacture of fabricated metal products, except machinery and equipment |
| 29 | Manufacture of machinery and equipment n.e.c. |
| 30 | Manufacture of office machinery and computers |
| 31 | Manufacture of electrical machinery and apparatus n.e.c. |
| 32 | Manufacture of radio, television and communication equipment and apparatus |
| 33 | Manufacture of medical, precision and optical instruments, watches and clocks |
| 34 | Manufacture of motor vehicles, trailers and semi-trailers |
| 35 | Manufacture of other transport equipment |
| 36 | Manufacture of furniture; manufacturing n.e.c. |
| 72 | Computer and related service activities |
| 73 | Research and development services |

Source: Classification of Economic Activities NACE Rev. 1.1 (Commission Regulation [EC] No 29/2002).